\documentclass[11pt,a4paper]{article}
\pdfoutput=1
\usepackage{geometry}
\usepackage[pdftex]{graphicx}
\usepackage{color}
\usepackage{framed}
\usepackage{amssymb}
\usepackage{latexsym}
\usepackage{amsmath}
\usepackage{enumitem}
\usepackage{bm}
\renewcommand{\Re}{\text{Re}}
\renewcommand{\Im}{\text{Im}}
\usepackage[width=0.9\textwidth]{caption}
\usepackage[bookmarks=true,pdfborder={0 0 0}]{hyperref}
\usepackage[all]{hypcap}
\title{Scaling limits of periodic monopoles}
\author{Rafael Maldonado\footnote{{\tt rafael.maldonado@durham.ac.uk}}\vspace{0.2cm}\\\emph{Department of 
Mathematical Sciences,}\\\emph{South Road, Durham DH1 3LE, UK}}

\begin{document}
\maketitle
\begin{abstract}
\noindent The purpose of this note is to explore the structure of singly periodic monopoles for different 
values of the size to period ratio.  The transition between a chain of small monopoles and the approximately 
two dimensional chain of large monopoles takes us through a region with an unintuitive dependence on the 
periodic direction.  The focus is mainly on the smooth SU(2) monopole of charge 2.
\end{abstract}
\section{Introduction}\label{introduction}
Periodic monopoles have been studied in the past by means of the generalised Nahm transform.  In 
\cite{ChK01}, it was shown that the Bogomolny (monopole) equations on $\mathbb{R}^2\times S^1$ with 
coordinates 
$\zeta=x+\text{i}y$ and $z\sim z+\beta$ are mapped to solutions of Hitchin equations
\begin{equation*}
F_{s\bar{s}}\,=\,-\tfrac{1}{4}[\Phi,\Phi^\dag]\qquad\qquad D_{\bar{s}}\Phi\,=\,0
\end{equation*}
on a cylinder described by $(r,t)\in\mathbb{R}\times S^1$, with $s=r+\text{i}t$.  We will work with the 
notation of \cite{HW09}, in which a class of solutions to the Hitchin equations corresponding to a chain of 
charge $2$ is
\begin{equation*}
\Phi\,=\,\begin{pmatrix}0&\mu_+\text{e}^{\psi/2}\\\mu_-\text{e}^{-\psi/2}&0\end{pmatrix}\qquad\qquad A_{\bar{s}}\,=\,a\sigma_3\qquad\qquad A_s\,=\,-\bar{a}\sigma_3\label{pg}
\end{equation*}
where the characteristic equation of $\Phi$ is fixed by the spectral curve as described in 
\cite{ChK01},\footnote{Here we follow the conventions of \cite{HW09,Mal13} for the sign of $K$.  The opposite 
sign was used in \cite{MW}, the only difference being a rotation of the monopole system by $\pi/2$ about the 
$z$ axis.}
\begin{equation*}
-\text{det}(\Phi)\,=\,\mu_+\mu_-\,=\,C\cosh(\beta s)+K/2
\end{equation*}
and $a$ and $\psi$ are functions of $(r,t)$ satisfying $4a=-\partial_{\bar{s}}\psi$,
\begin{equation}
\nabla^2\Re(\psi)\,=\,2\left(|\mu_+|^2\text{e}^{\Re(\psi)}-|\mu_-|^2\text{e}^{-\Re(\psi)}\right)\label{psieq}
\end{equation}
with $\Im(\psi)$ chosen in such a way that $\Phi$ has the correct $t$-period.
\par The inverse Nahm operator
\begin{equation}
\Delta\Psi\,=\,\begin{pmatrix}\bm{1}_2\otimes(2\partial_{\bar{s}}-z)+2A_{\bar{s}}&\bm{1}_2\otimes\zeta-\Phi\\\bm{1}_2\otimes\bar{\zeta}-\Phi^\dag&\bm{1}_2\otimes(2\partial_s+z)+2A_s\end{pmatrix}\,\Psi\,=\,0\label{invnahm}
\end{equation}
with solutions normalised to
\begin{equation*}
\int_{-\infty}^\infty dr\int_{-\pi/\beta}^{\pi/\beta}dt\,(\Psi^\dag\Psi)\,=\,\mathbf{1}_2
\end{equation*}
then allows a numerical construction of the monopole fields
\begin{equation*}
\hat{\Phi}\,=\,\text{i}\int_{-\infty}^\infty dr\int_{-\pi/\beta}^{\pi/\beta}dt\,(r\Psi^\dag\Psi)\qquad\qquad\hat{A}_i\,=\,\int_{-\infty}^\infty dr\int_{-\pi/\beta}^{\pi/\beta}dt\,(\Psi^\dag\partial_i\Psi)
\end{equation*}
and an analytical study of their symmetries.
\par We will focus on the two classes of solution with $\alpha=0$, \cite{Mal13}, for which the zeros of 
$\text{det}(\Phi)$ are placed either in the same or different entries of $\Phi$:
\begin{itemize}
\item The `zeros together' solution has $\Im(\psi)=0$ and
\begin{equation*}
\mu_+\,=\,C\cosh(\beta s)+K/2\qquad\qquad\mu_-\,=\,1.
\end{equation*}
\item The `zeros apart' solution has $\Im(\psi)=-\beta t$ and
\begin{equation}
\mu_\pm\,=\,\sqrt{C/2}\left(\text{e}^{\beta s/2}+W^{\pm1}\text{e}^{-\beta s/2}\right)\quad\text{where}\quad K/C\,=\,W+W^{-1}.\label{za}
\end{equation}
\end{itemize}
The choice of solution affects the symmetries and scattering processes of the corresponding monopoles, as 
will be discussed in section \ref{symmetries}.
\par In the monopole picture, these solutions correspond, for the complex modulus $|K|\gg2C$, to monopoles 
located approximately at $\zeta=\pm\sqrt{K/2}$ and with zero $z$-offset, \cite{Mal13}.  The positive real 
parameter $C$ determines the size to period ratio of the constituent monopoles.  Periodic monopoles have also 
been observed to split into two constituent energy peaks, in this case separated by $|C\sqrt{2/K}|$ when 
$|K|\gg2C$.\footnote{Such behaviour has been extensively researched for periodic instantons, which have 
monopole constituents \cite{LL98}.  The case of periodic monopoles displays an important difference with the 
caloron case, in that constituents are always present due to the asymptotic holonomy always being non-trivial 
\cite{Mal13}, although their separation reduces as $C\to0$.  Furthemore, it is not yet clear whether there is 
in this case a meaningful description of the constituents as separate entities in their own right.}  The aim 
of this paper is to study how the monopole shape and location depend on $K$ and $C$.
\par Periodic monopoles of charge $1$ and $2$ were constructed in \cite{War05} and \cite{HW09}.  In 
\cite{Mal13}, the `spectral approximation' was introduced, allowing the monopole fields to be read off from 
the spectral curve when $C$ is large.  This paper gives further evidence on the validity of this 
approximation and illustrates how it arises from the well separated monopoles at small $C$.  The effect of 
$C$ on the moduli space of solutions was discussed further in \cite{MW}.
\par This paper is arranged as follows.  In section \ref{symmetries} the symmetries of charge $2$ periodic 
monopoles are given.  The limits of small and large $C$ are discussed in sections \ref{smallC} and \ref{largeC}.  
The intermediate regime is then studied in section \ref{intermediateC}, and some comments on the generalisation 
to higher charges are made in section \ref{highercharges}.
\section{Monopole symmetries}\label{symmetries}
Spatial symmetries of the monopole can be deduced from the structure of the inverse Nahm operator 
\eqref{invnahm} as follows.  The symmetries of the spectral curve motivate a transformation of the 
coordinates on the Hitchin cylinder, such that the transformed Nahm/Hitchin fields can be expressed in terms 
of a gauge transformation of the original fields.  A suitable transformation of the spatial $(\zeta,z)$ 
coordinates which twist the inverse Nahm operator leaves its kernel unchanged, implying the monopole fields 
at the transformed spatial coordinates are gauge equivalent to the original monopole fields.  This procedure 
was discussed in \cite{HW09} and extended in \cite{Mal13}, and the reader is referred to these for a detailed 
discussion.  This section summarises the symmetries relevant to the charge $2$ periodic monopole, extending 
those presented in \cite{Mal13} by considering more carefully the branching structure of the solutions for 
$|K|\leq2C$.
\par In the `zeros together' case the monopole fields display the spatial symmetries 
$(\zeta,z)\sim(\bar{\zeta},-z)$ for $K\in\mathbb{R}$, and $(\zeta,z)\sim(\text{i}\bar{\zeta},-z)$ for 
$K\in\text{i}\mathbb{R}$.  In both cases $(\zeta,z;K)\sim(\text{i}\zeta,z;-K)$, such that when $K=0$ there is 
an enhanced $C_4$ symmetry $(\zeta,z)\sim(\text{i}\zeta,z)$.
\par The `zeros apart' case enjoys the symmetry $(\zeta,z)\sim(\bar{\zeta},-z)$ for $W\in\mathbb{R}$ and 
$(\zeta,z)\sim(\text{i}\bar{\zeta},-z)$ for $W\in\text{i}\mathbb{R}$, in both cases with 
$(\zeta,z,W)\sim(\bar{\zeta},\beta/2-z,\overline{W}^{-1})$ relating the incoming and outgoing legs of the 
geodesics on the half-lines $W=p^2$ and $W=\text{i}p^2$ for $p>0$.  There are two cases with enhanced 
symmetry, which will be of particular interest in this paper:
\begin{itemize}
\item for $W=1$, $(\zeta,z)\sim(\zeta,\beta/2-z)\sim(\zeta,z+\beta/2)$,
\item for $W=\text{i}$, $(\zeta,z)\sim(\text{i}\zeta,\beta/2-z)\sim(\text{i}\zeta,z+\beta/2)$.
\end{itemize}
These symmetries involve a chain in the sign of $\hat{\Phi}$, but correctly describe symmetries of the energy 
density.  Unlike for the charge $2$ monopole in $\mathbb{R}^3$, the symmetries are always discrete due to the 
splitting into constituents.  As $|C|$ is reduced, the constituents move closer together and in the limit 
$C\to0$ the phase of $C$ can have no effect, reproducing the axially symmetric charge $2$ monopole 
\cite{MS07} in this limit.
\section{Small C}\label{smallC}
\begin{figure}
\centering
\vspace{-0.3cm}
\includegraphics[width=0.5\linewidth]{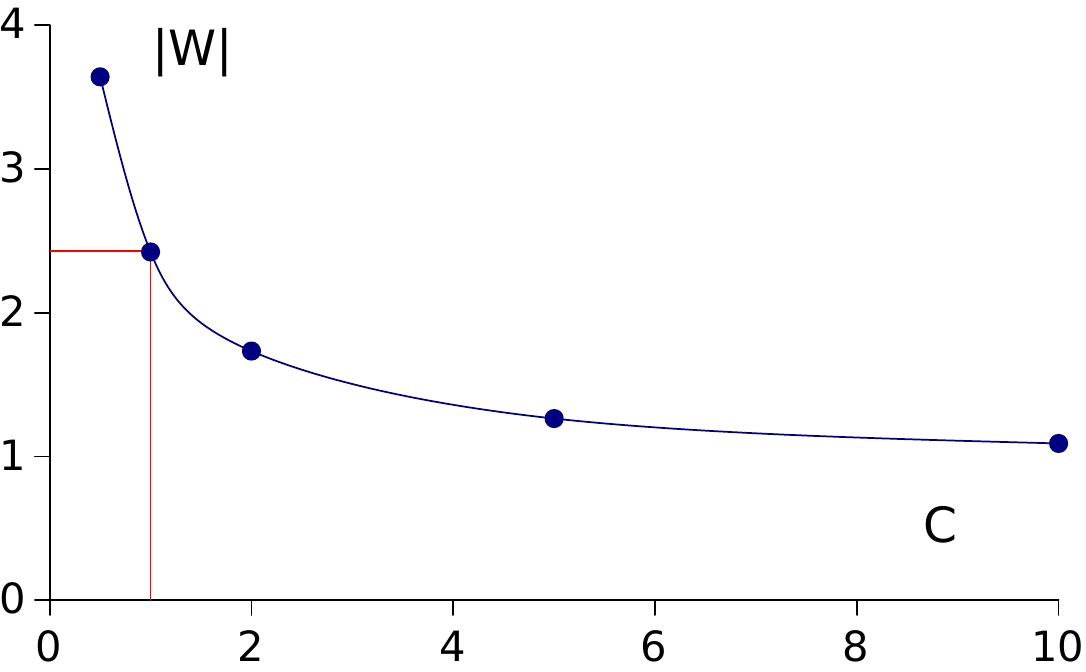}
\caption{$|W|$ against $C$, showing how the value of $|W|$ at which the monopole Higgs zeros coincide in the 
`zeros apart' configuration depends on $C$, both for $W\in\mathbb{R}$ and $W\in\text{i}\mathbb{R}$.  For 
`zeros together', the monopole zeros always coincide when $K=0$.}\label{Crunning}
\end{figure}

In the limit of small size to period ratio monopole chains behave like monopoles in $\mathbb{R}^3$, whose 
energy density peaks roughly at the location of the zeros of the Higgs field.  The two scattering processes 
identified in section \ref{symmetries} correspond in this limit to the Atiyah-Hitchin rounded cone (`zeros 
together') and trumpet (`zeros apart'), as geodesic submanifolds of the full four dimensional moduli space, 
\cite{MW}.  Although it is straightforward to reach the above conclusions numerically, the limit is 
nevertheless delicate to provide analytically in the present formulation.  In particular, it is not clear how 
the ALG type metric reduces to the usual ALF of monopoles in $\mathbb{R}^3$ \cite{ChK02}.  In this limit we 
also see that the coordinate $W$ goes bad, in the sense that the value of $W$ at which the monopole Higgs 
zeros coincide increases as $C\to0$, as shown in figure \ref{Crunning}.
\par
\begin{figure}
\begin{minipage}{0.485\linewidth}
\centering
\vspace{-0.3cm}
\includegraphics[width=0.95\linewidth]{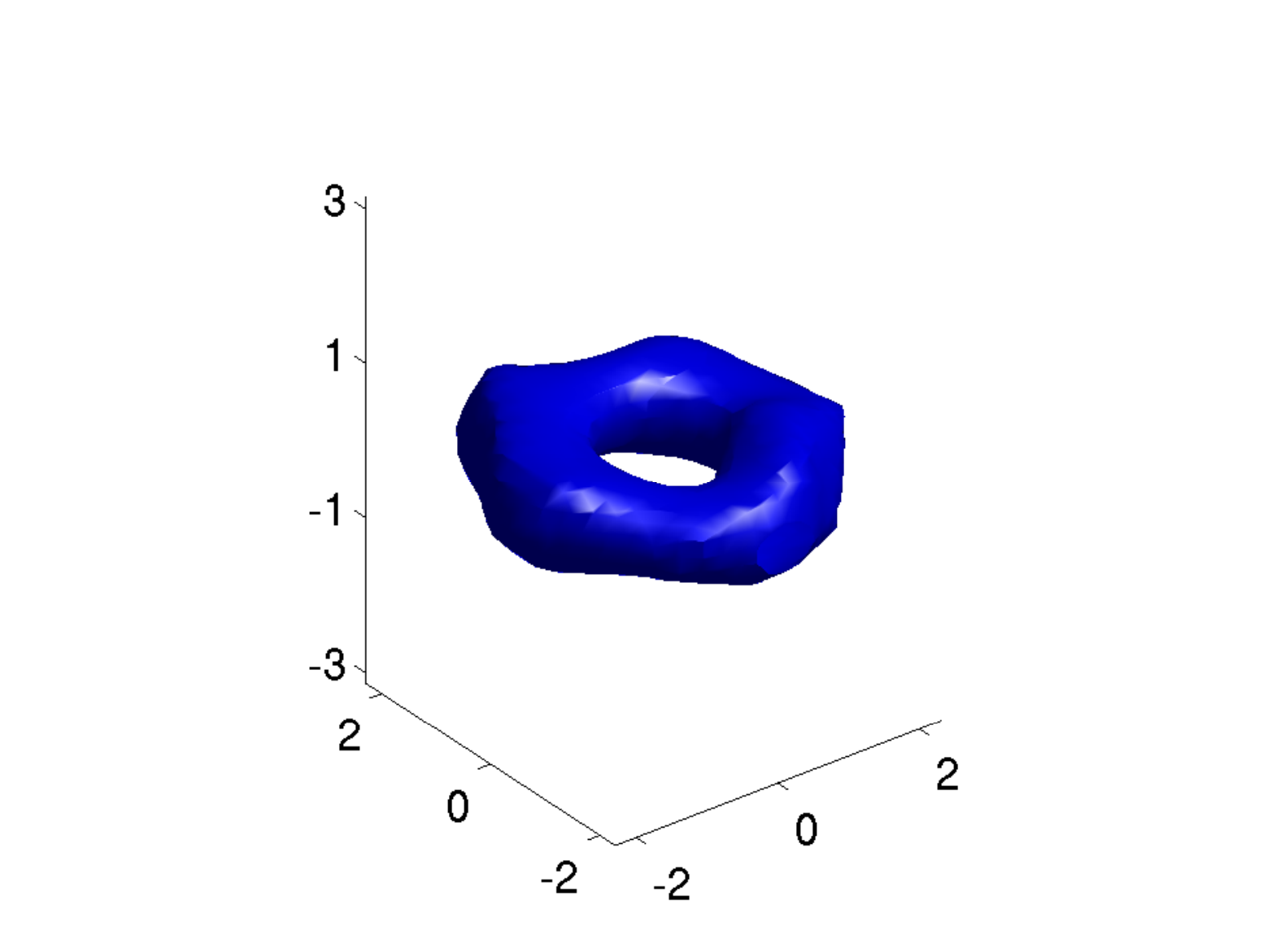}
\end{minipage}
\begin{minipage}{0.485\linewidth}
\centering
\includegraphics[width=0.95\linewidth]{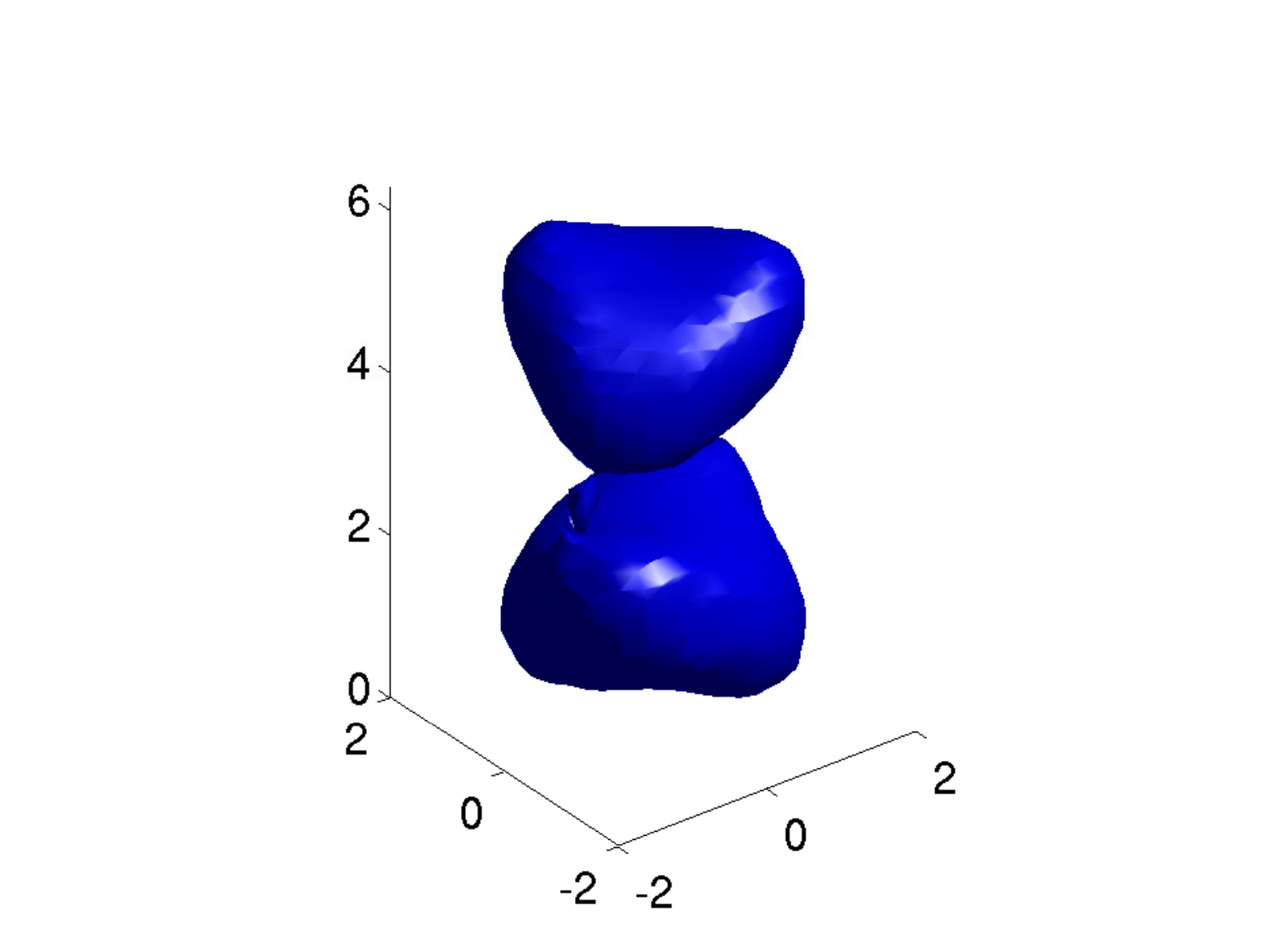}
\end{minipage}
\vspace{-0.5cm}
\caption{Energy density for charge $2$ monopole in the `zeros together' (left) and `zeros apart' (right) 
configuration with $C=1$ and $W=\text{i}$.}\label{C1K0}
\end{figure}

\begin{figure}
\begin{minipage}{0.485\linewidth}
\centering
\vspace{-0.3cm}
\includegraphics[width=0.95\linewidth]{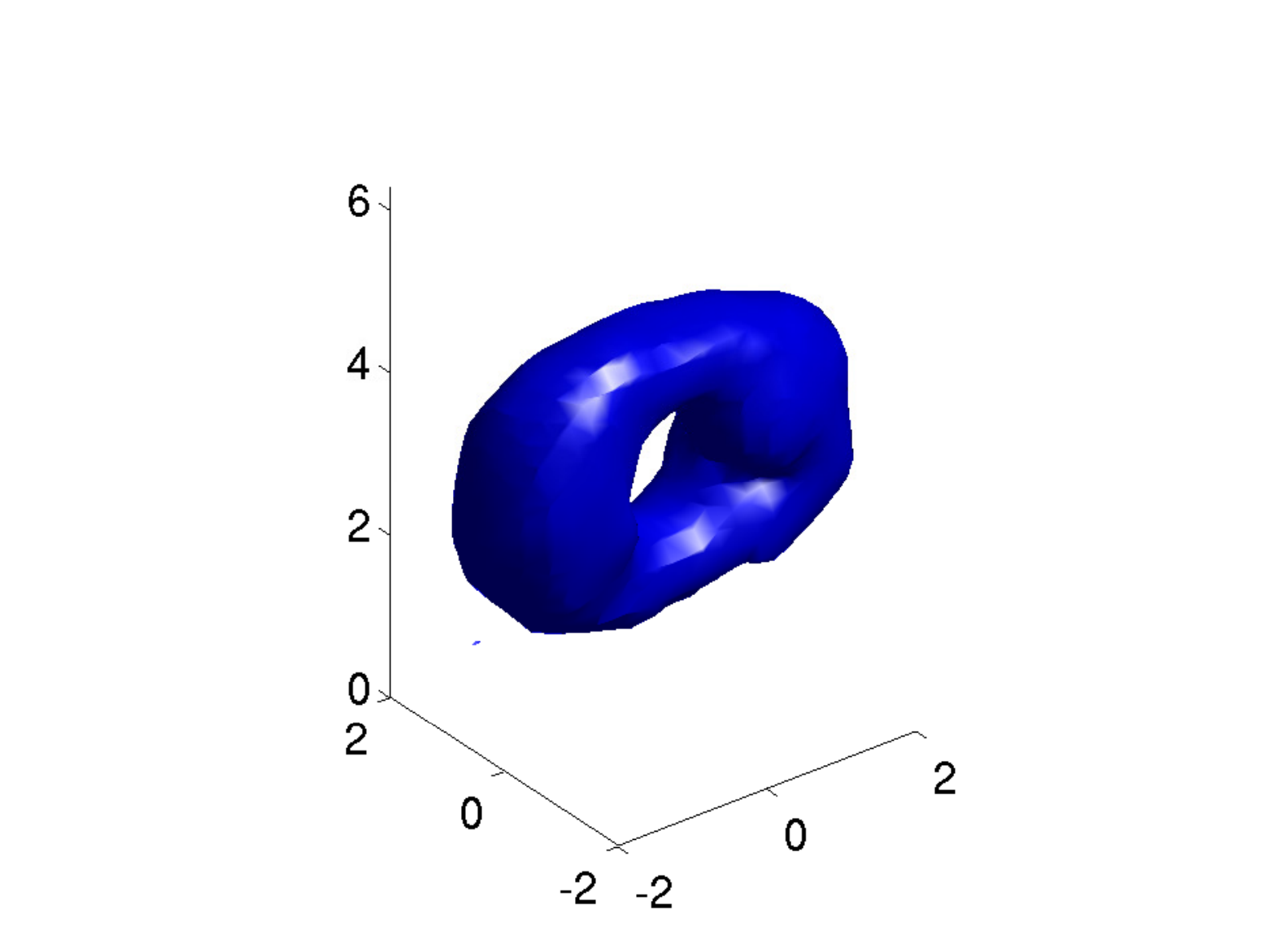}
\end{minipage}
\begin{minipage}{0.485\linewidth}
\centering
\includegraphics[width=0.95\linewidth]{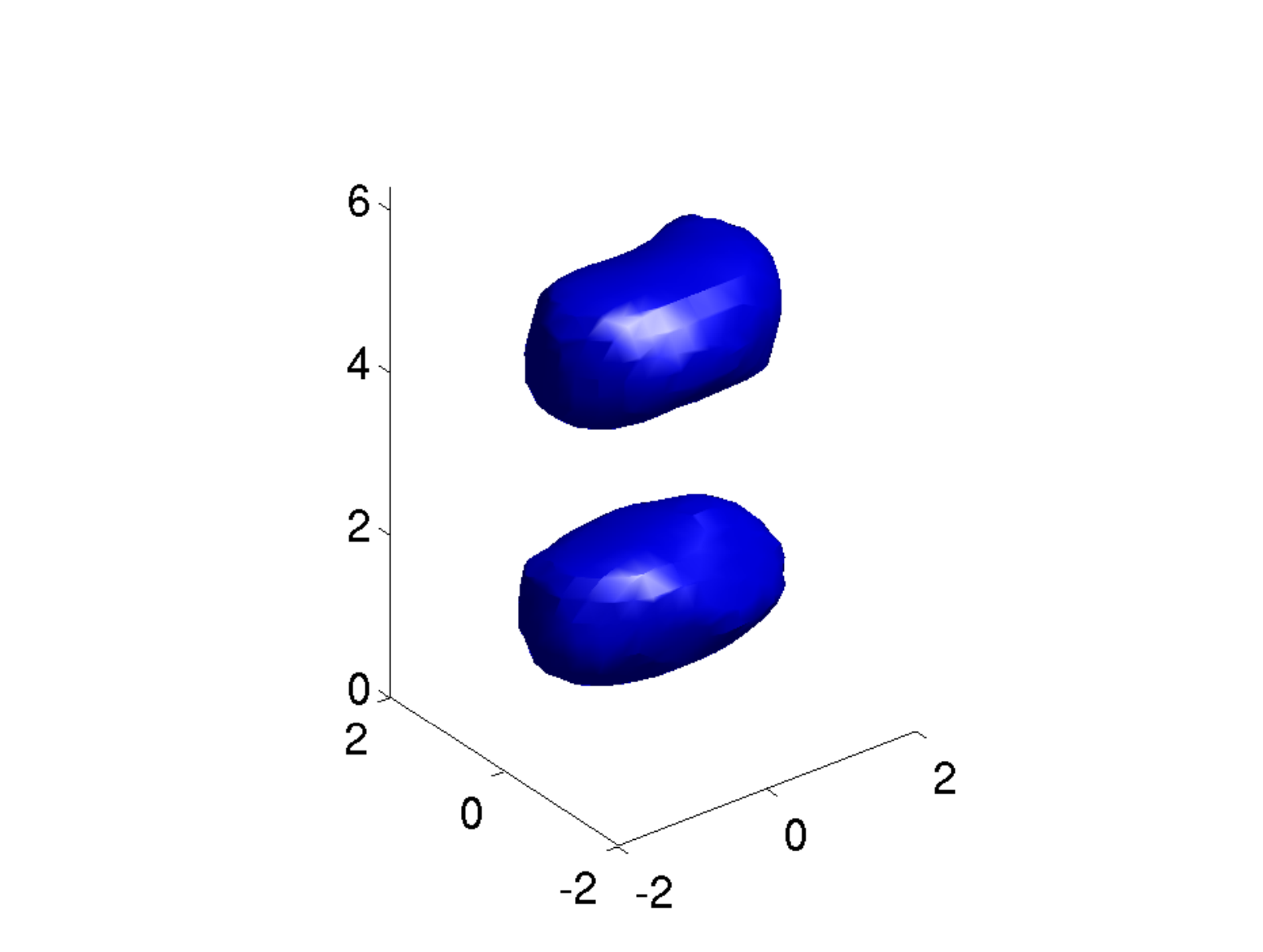}
\end{minipage}
\vspace{-0.5cm}
\caption{Energy density for charge $2$ monopole in the `zeros apart' configuration with $C=1$ and $W=2.43$ 
(left) and $W=1$ (right).}\label{C1K2}
\end{figure}

 The particularly symmetric case with $K=0$ is shown in figure \ref{C1K0} for the `zeros together' and 
`zeros apart' solutions, displaying the expected spatial symmetries (section \ref{symmetries} and 
ref.~\cite{Mal13}).  The `zeros apart' geodesic for $C=1$ and $W>1$ has two monopole chains incoming along 
the $x$ axis, whose energy density is peaked at the Higgs zeros.  At $W\approx2.43$ (figure \ref{Crunning}) 
the Higgs zeros coincide to give a toroidal configuration (figure \ref{C1K2} left).  Reducing $K$ further, the 
ring breaks up along the $z$ axis, giving two copies of a charge $1$ monopole when $W=1$ (figure \ref{C1K2} 
right, see also \cite{HW09,Mal}), which move apart parallel to the $x$ axis for $W<1$.  The geodesic with 
$W\in\text{i}\mathbb{R}$ again involves a double scattering, though this time the `doubled' charge $1$ chain 
is not encountered and chains depart at $90^\circ$ to the incoming chains.
\subsection{Regaining the Nahm equations}
Defining the combinations $\Phi=\text{i}(T_1+\text{i}T_2)$, $A_r=T_0$, $A_t=T_3$ of the Hitchin fields with 
$T_i=\frac{1}{2}\text{i}f_i\sigma_i$, we take the limit $C=0$, such that $\text{det}(\Phi)=-K/2$ and the 
Hitchin fields on the cylinder depend only on $r$.  This reproduces the usual Nahm equations in 
$\mathbb{R}^3$, and although this approach is only valid in the strict limit $C\to0$, it is interesting to 
note how the different `zeros together' and `zeros apart' solutions can still be seen in this limit.
\par In the above notation, the Hitchin equations become Nahm equations, such that the functions $f_i$ satisfy
\begin{equation}
\frac{df_i}{dr}\,=\,\frac{1}{2}\,\epsilon_{ijk}f_jf_k\label{Nahm}
\end{equation}
and the Hitchin fields become
\begin{equation*}
\Phi\,=\,-\frac{1}{2}\begin{pmatrix}0&f_1+f_2\\f_1-f_2&0\end{pmatrix}\qquad\qquad\frac{d\psi}{dr}\,=\,2f_3
\end{equation*}
where we have chosen a gauge with $A_r=0$.  The spectral curve tells us that
\begin{equation*}
-\text{det}(\Phi)\,=\,\tfrac{1}{4}(f_1^2-f_2^2)\,=\,C\cosh(\beta s)+K/2,
\end{equation*}
and \eqref{Nahm} immediately requires $C=0$.  In this form, with $\alpha=0$ and $K\in\mathbb{R}$, the Nahm 
equations can easily be solved in terms of elliptic functions \cite{BPP82,MS07}.
\par For real $f_i$ the Nahm transform provides a clear link between the symmetries of $(\zeta,z)$ and those 
of $(\Phi,T_3)$.  It is thus expected that there will be different solutions to the Nahm equations 
corresponding to the relative magnitudes of $f_1^2,f_2^2,f_3^2$.  We note from \cite{BPP82,MS07} that for 
large $K$ the monopoles are localised along the axis $e_i$ corresponding to the largest of the $f_i^2$.  We 
will fix $\zeta=e_1+\text{i}e_2$ and $z=e_3$ (this is a gauge choice on the Nahm data), with monopoles 
incoming along $e_1$.
\\ \\
First of all we take $f_1^2\geq f_2^2\geq f_3^2$, and define a function $a(K)$ and the elliptic modulus 
$k\in[0,1]$ by
\begin{equation*}
f_1^2-f_2^2\,=\,2K\qquad\qquad f_1^2-f_3^2\,=\,a^2\qquad\qquad2K\,=\,a^2k^2
\end{equation*}
which are solved in terms of Jacobi elliptic functions defined for $|ar|<\text{\textbf{K}}(k)$,
\begin{equation*}
f_1\,=\,a\text{dc}_k(ar)\qquad\qquad f_2\,=\,ak'\text{nc}_k(ar)\qquad\qquad f_3\,=\,ak'\text{sc}_k(ar).
\end{equation*}
In the limit $K\to0$ the monopole chains approach one another and\footnote{This expression for $\psi$ can also 
be obtained from the `zeros together' Hitchin equations with $K=0$ by use of an approximate solution 
\cite{Har} which improves as $\beta\to\infty$.  This approach clarifies the transition, as $\beta$ is 
increased, between the smooth Nahm/Hitchin data valid on the entire length of the cylinder and the solution 
valid on the finite interval $ar\in\pi/2(-1,1)$ with singularities at the endpoints (note that 
$\text{\textbf{K}}(0)=\pi/2$).  In essence, the singular solution for $\psi$ arises from the $\beta\to\infty$ limit of 
the boundary conditions $\psi\sim\pm\beta r$ for $r\to\pm\infty$.  It is pleasing to see that the limits 
$\beta\to\infty$ and $C\to0$ yield the same result, in both cases describing chains of small monopoles.}
\begin{equation*}
f_1\,=\,f_2\,=\,a\sec(ar)\qquad\qquad f_3\,=\,a\tan(ar)\qquad\qquad\psi\,=\,2\log(2ab\sec(ar))
\end{equation*}
for some constant $b$.
\par The equality $f_1=f_2$ in this limit describes a monopole configuration which is axially symmetric about 
the periodic axis, and leads to $90^\circ$ scattering in the plane when $K$ becomes negative (when 
$f_2^2\geq f_1^2\geq f_3^2$).
\par
\begin{figure}
\begin{minipage}{0.485\linewidth}
\centering
\includegraphics[width=0.85\linewidth]{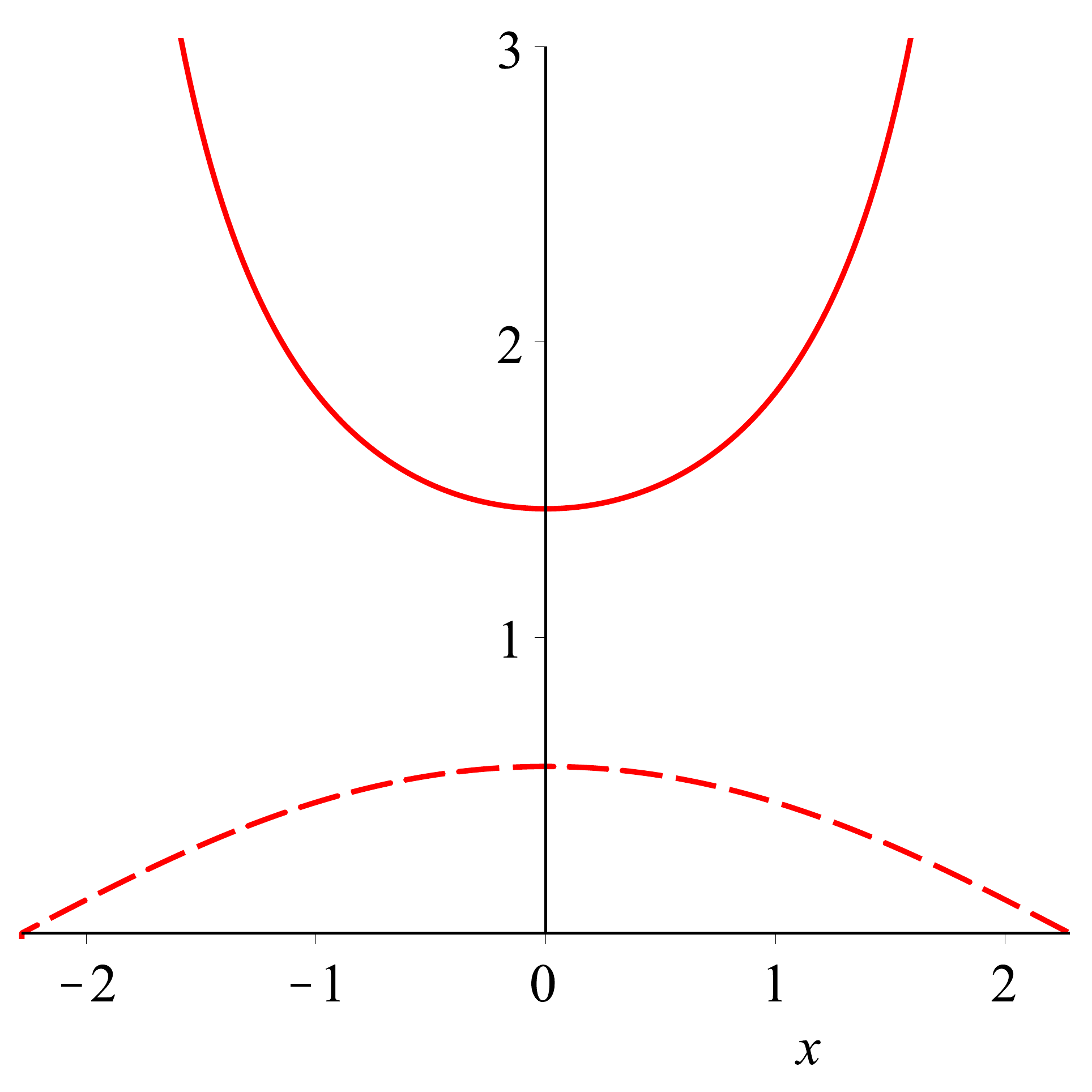}
\end{minipage}
\begin{minipage}{0.485\linewidth}
\centering
\includegraphics[width=0.85\linewidth]{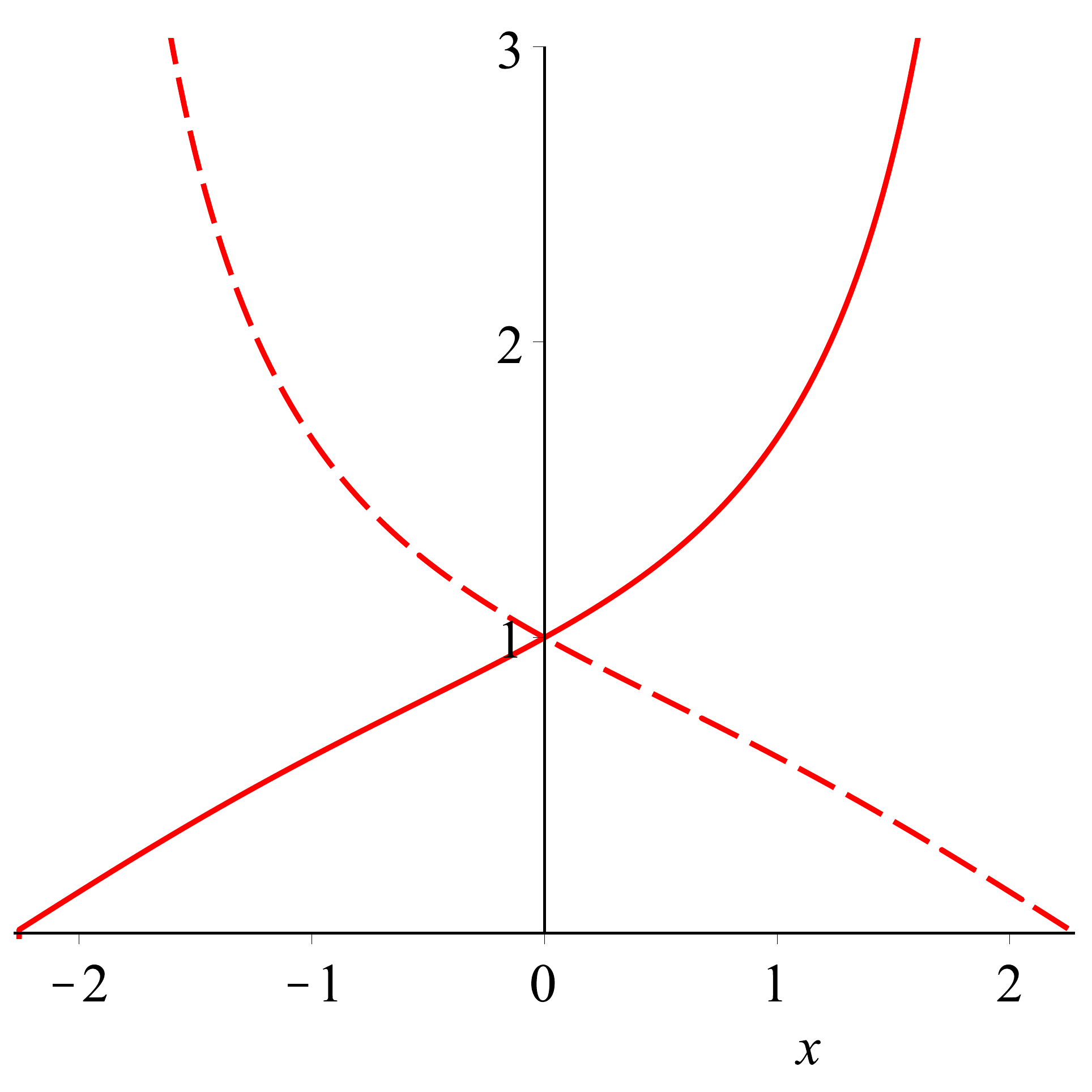}
\end{minipage}
\caption{$(f_1+f_2)/a$ (solid) and $(f_1-f_2)/a$ (dashed) for $k=0.9$ plotted against $x=ar$ (left, for 
$f_1^2>f_2^2>f_3^2$) and $x=\sqrt{2K}r$ (right, for $f_1^2>f_3^2>f_2^2$).}\label{zta}
\end{figure}

 Fig.~\ref{zta} (left) shows a plot of $f_1\pm f_2$ for $k=0.9$, illustrating how both zeros are in 
the same component of $\Phi$ (i.e. the `zeros together' solution).
\\ \\
On the other hand, there is the possibility of having $f_1^2\geq f_3^2\geq f_2^2$.  This time,
\begin{equation*}
f_1^2-f_2^2\,=\,2K\qquad\qquad f_1^2-f_3^2\,=\,a^2\qquad\qquad2Kk^2\,=\,a^2
\end{equation*}
and the solution is
\begin{equation*}
f_1\,=\,\sqrt{2K}\text{dc}_k(\sqrt{2K}r),\qquad f_2\,=\,\sqrt{2K}k'\text{sc}_k(\sqrt{2K}r),\qquad f_3\,=\,\sqrt{2K}k'\text{nc}_k(\sqrt{2K}r).
\end{equation*}
This time, when $K=0$ we simply have $f_1=f_2=f_3=0$, which is the Nahm data for a single monopole.  
Fig.~\ref{zta} (right) shows $f_1\pm f_2$.  The zeros of $\Phi$ are now in different components and 
scattering is consistent with the `zeros apart' solution.
\section{Large C}\label{largeC}
\begin{figure}
\centering
\includegraphics[width=\linewidth]{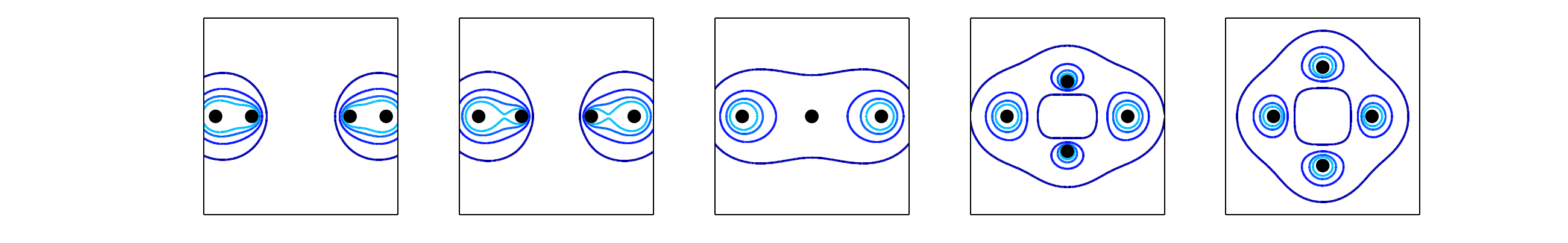}
\caption{Sequence (to be read from left to right and top to bottom) showing the scattering of monopole 
chains with $K\in\mathbb{R}$ within the spectral approximation.}\label{fig04a}
\end{figure}

In the opposite limit, of large monopole size to period ratio, the structure again simplifies.  As $C$ is 
increased, the fields become increasingly independent of $z$ and the spectral approximation \cite{Mal13} 
becomes an accurate description of the monopole.  The monopole Higgs field is known explicitly in this limit 
and can be read off directly from the spectral curve,
\begin{equation*}
\hat{\Phi}\,=\,\frac{\text{i}}{\beta}\,\Re\left(\cosh^{-1}\left(\frac{2\zeta^2-K}{2C}\right)\right)\sigma_3,
\end{equation*}
and the energy density calculated through
\begin{equation}
\mathcal{E}\,=\,\frac{1}{4}\,\nabla^2|\text{tr}(\hat{\Phi}^2)|.\label{endens}
\end{equation}
Geodesic motion with $K\in\mathbb{R}$ describes the movement of four lumps of energy density located at 
$\zeta=\pm\sqrt{K/2\pm C}$ undergoing a double scattering via a cross-shaped configuration at $K=0$, as shown 
in figure \ref{fig04a}.
\par In the large $C$ limit there is also a simplification in the solutions of the Hitchin equations.  The 
$C$ dependence of $\mu_\pm$ in \eqref{psieq} means a non-trivial solution for $\nabla^2\Re(\psi)$ is only 
supported at small $C$ and in the vicinity of the two regions $\mu_\pm\approx0$.  Thus, in the large $C$ 
limit, the smooth solution to \eqref{psieq} approaches the singular solution obtained by setting both sides 
to zero.
\par Details of the geometry of the moduli space in this limit are given in \cite{Mal13,MW}.  The solutions to 
the Nahm/Hitchin equations in this limit, referred to above, imply that the metric obtained from this data 
depends only on $\text{det}(\Phi)$, and is hence the same for the `zeros together' and `zeros apart' 
solutions.  This is identical to the metric found from the spectral approximation to the monopole fields in 
\cite{Mal13}.  This metric has two logarithmic singularities on the $K$ plane.  Numerically, for $C\gtrsim36$ 
we recover the $z$-independent fields as approximately cylindrical tubes of energy density aligned with the 
$z$ axis.
\subsection{Nahm transform for large C}
The large scale limit allows a demonstration of an example of the Nahm transform for the construction of 
solutions to the Hitchin equations on $\mathbb{R}^2$ (twice dimensionally reduced self-dual Yang-Mills 
equations).  The general theory of Nahm transforms \cite{Jar04} suggests that the Nahm transform on 
$\mathbb{R}^2$ is `self-reciprocal' \cite{CG84}, thus mapping the large $C$ limit of the periodic monopole to 
Hitchin equations on $\mathbb{R}^2$, with a different topology and boundary conditions.  It is not clear how 
to `unwrap' the Hitchin cylinder in this limit, or how one might deal with the singular nature of the 
solutions.  However, as a step towards understanding this instance of the Nahm transform, we show that in 
this limit the `spectral approximation' can also be applied to the forward Nahm transform, allowing us to 
reproduce the initial Nahm data from the approximate monopole fields.  Below we look specifically at the 
charge $1$ periodic monopole, although the argument can equally be applied to higher charges.
\\ \\
The inverse Nahm operator for the charge $1$ periodic monopole \cite{War05} is
\begin{equation}
\Delta\Psi\,=\,\begin{pmatrix}2\partial_{\bar{s}}-z&\zeta-\Phi\\\bar{\zeta}-\Phi^\dag&2\partial_s+z\end{pmatrix}\begin{pmatrix}\psi_{11}&\psi_{12}\\\psi_{21}&\psi_{22}\end{pmatrix}\,=\,0,\label{ch1invnahm}
\end{equation}
where $\Phi=C\cosh(\beta s)$.  To study the large $C$ limit, we suppress $z$ dependence (setting $z=0$ above) 
and define new fields
\begin{equation*}
\text{i}\int_{\mathbb{R}^2}dr\,dt\,(s\Psi^\dag\Psi)\,=\,\text{i}\hat{\phi}\qquad\qquad\int_{\mathbb{R}^2}dr\,dt\,(\Psi^\dag\partial_j\Psi)\,=\,\hat{a}_j
\end{equation*}
where $\text{i}\hat{\phi}=\hat{\Phi}-\text{i}\hat{A}_z$ and $j=x,y$.  These fields satisfy Hitchin equations 
in $\mathbb{R}^2$.  Equation \eqref{ch1invnahm} has an approximate solution valid at large $C$, 
\cite{War05,Mal13}, in which the columns of $\Psi$ are Gaussian peaks at $s=\pm s_0$ with $s_0(\zeta)$ defined 
through $C\cosh(\beta s_0)=\zeta$.  In this limit the monopole fields are $\hat{\phi}=s_0\sigma_3$, 
$\hat{a}=0$.
\par The idea is to use these approximate monopole fields to explicitly perform the forward Nahm transform, 
i.e. starting from $\hat{\phi},\hat{a}$ obtain $\Phi$ and $A$.  The Nahm operator is
\begin{equation}
\hat{\Delta}\hat{\Psi}\,=\,\begin{pmatrix}2\partial_{\bar{\zeta}}&0&s-s_0&0\\0&2\partial_{\bar{\zeta}}&0&s+s_0\\\bar{s}-\bar{s}_0&0&2\partial_\zeta&0\\0&\bar{s}+\bar{s}_0&0&2\partial_\zeta\end{pmatrix}\begin{pmatrix}\hat{\psi}_1\\\hat{\psi}_2\\\hat{\psi}_3\\\hat{\psi}_4\end{pmatrix}\,=\,0,\label{ch1forwardnahm}
\end{equation}
normalised solutions to which should give the charge $1$ Nahm/Hitchin data,
\begin{equation*}
\Phi\,=\,\int_{\mathbb{R}^2}dx\,dy\,\zeta\sum_{i=1}^4|\hat{\psi}_i|^2\,=\,C\cosh(\beta s)\qquad\qquad A\,=\,\int_{\mathbb{R}^2}dx\,dy\,\sum_{i=1}^4\bar{\hat{\psi}}_i\partial_j\hat{\psi}_i\,=\,0.
\end{equation*}
Solutions to the forward Nahm operator \eqref{ch1forwardnahm} are found using the same ideas as those for the 
inverse transform.  First of all, we note that the equations for $\hat{\psi}_1$ and $\hat{\psi}_3$ decouple 
from those for $\hat{\psi}_2$ and $\hat{\psi}_4$.  Writing $\zeta_0=C\cosh(\beta s)$ and 
$\zeta=\zeta_0+\delta$, we have
\begin{equation*}
s-s_0\,=\,\frac{1}{\beta}\cosh^{-1}\left(\frac{\zeta_0}{C}\right)-\frac{1}{\beta}\cosh^{-1}\left(\frac{\zeta_0+\delta}{C}\right)\,=\,-\frac{\zeta-\zeta_0}{\beta}\xi+\mathcal{O}(\delta^2)
\end{equation*}
where $\xi^{-1}=C\sinh(\beta s)$.  $\hat{\psi}_1$ and $\hat{\psi}_3$ are supported away from $s=0$, and we 
make the Ansatz $\hat{\psi}_{1,3}\sim\text{exp}(-c|\zeta-\zeta_0|^2)$, resulting in $c=|\xi|/(2\beta)$ and 
$\hat{\psi}_3=-\xi^{-1/2}\bar{\xi}^{1/2}\hat{\psi}_1$.
\par The important point now is that, if we remain on the correct branch of $\cosh^{-1}$, the quantity 
$(s+s_0)$ will never be close to zero (as in \cite{War05}, we must avoid the points $\zeta_0=\pm C$).  Thus, 
$\hat{\psi}_{2,4}$ are small and slowly varying compared to $\hat{\psi}_{1,3}$.  We thus approximate 
$\hat{\psi}_{2,4}\approx0$, so that normalising gives
\begin{equation*}
|\hat{\psi}_{1,3}|^2\,=\,\frac{|\xi|}{2\pi\beta}\,\text{e}^{-|\xi||\zeta-\zeta_0|^2/\beta}.
\end{equation*}
The consistency relation $\hat{\psi}_3=\pm\bar{\hat{\psi}}_1$ fixes the phases on $\hat{\psi}_{1,3}$,
\begin{equation}
\hat{\psi}_1\,=\,-\left(\frac{\xi}{2\pi\beta}\right)^{1/2}\,\text{e}^{-|\xi||\zeta-\zeta_0|^2/(2\beta)},\qquad\qquad\hat{\psi}_3\,=\,\left(\frac{\bar{\xi}}{2\pi\beta}\right)^{1/2}\,\text{e}^{-|\xi||\zeta-\zeta_0|^2/(2\beta)},\label{psihat}
\end{equation}
yielding the expected Hitchin fields, $\Phi=\zeta_0=C\cosh(\beta s)$, $A=0$.  Note the solution \eqref{psihat} 
is again exponentially localised, and the scaling with $\beta$ in opposite to that of $\Psi$.
\section{Intermediate C}\label{intermediateC}
Now the small and large $C$ limits have been established, our aim is to understand the intermediate regime.  
Here we focus on the `zeros apart' case, which displays a rich $z$ behaviour.  It is instructive to study the 
geodesics $W\in\mathbb{R}$, $W\in\text{i}\mathbb{R}$ and $|W|=1$.
\par
\begin{figure}
\begin{minipage}{0.485\linewidth}
\centering
\vspace{-0.3cm}
\includegraphics[width=0.95\linewidth]{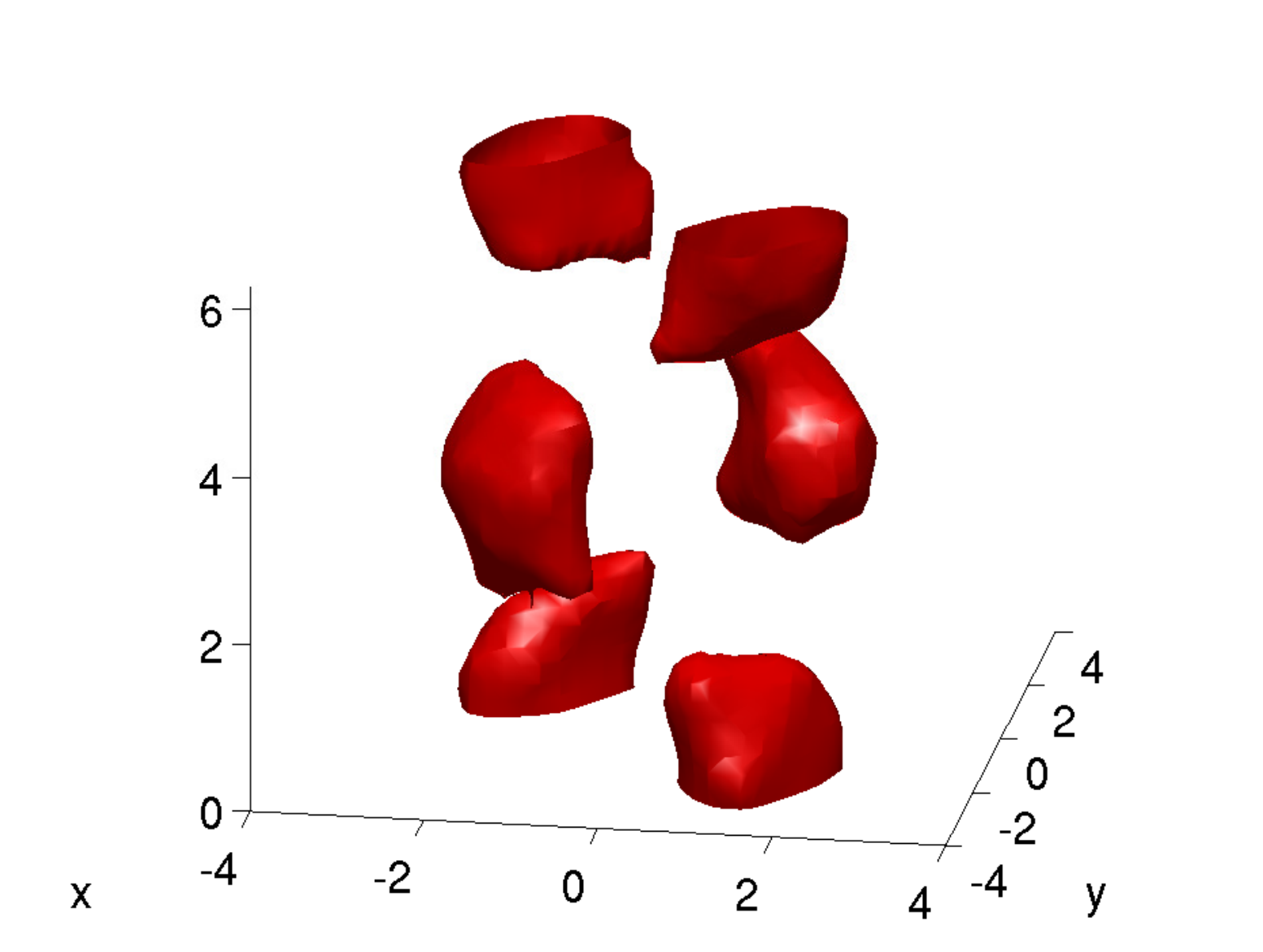}
\end{minipage}
\begin{minipage}{0.485\linewidth}
\centering
\includegraphics[width=0.95\linewidth]{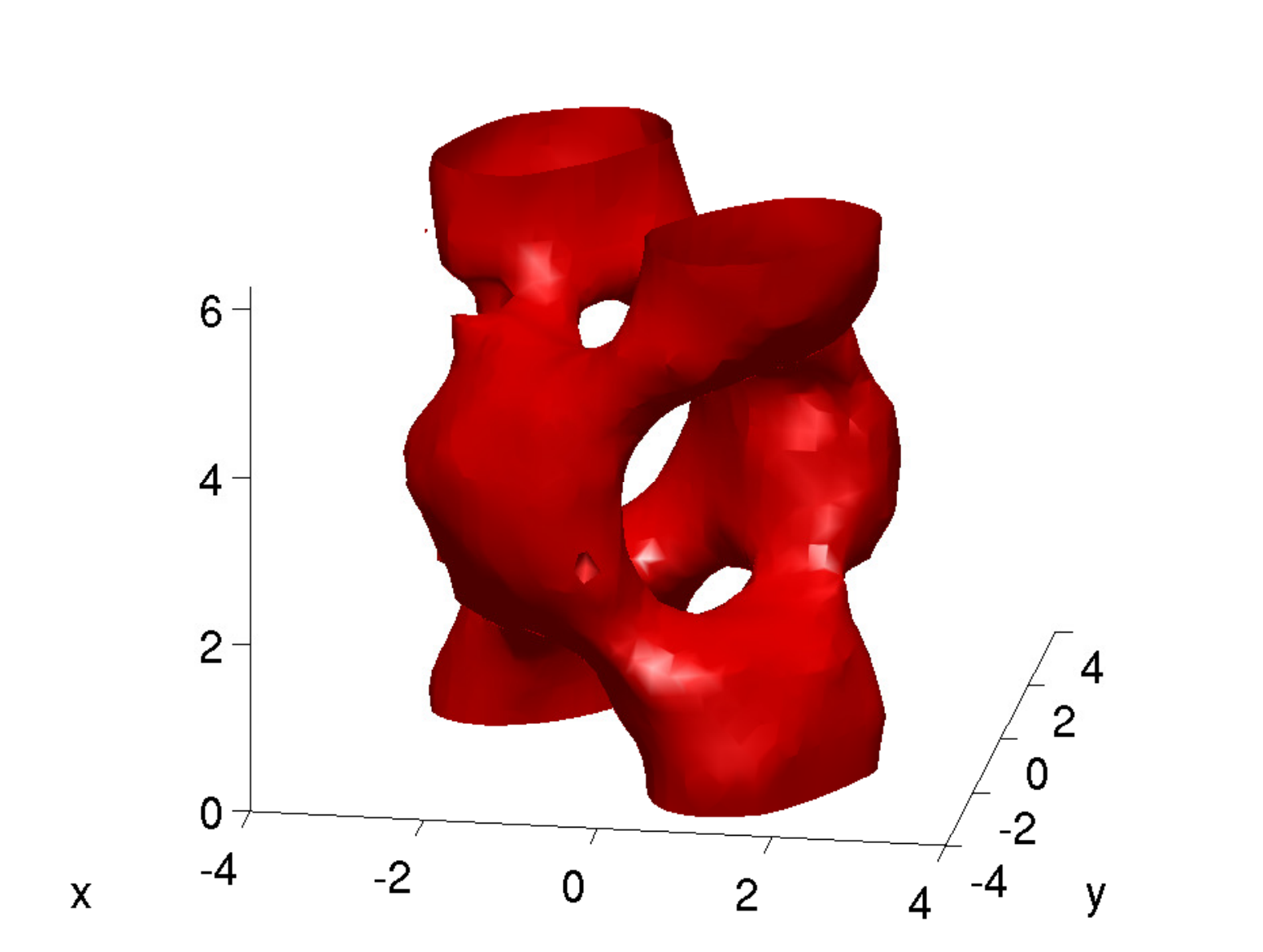}
\end{minipage}
\caption{Energy density isosurfaces for $C=4$, $W=\text{i}$.  On the left we see the constituent 
structure, and on the right the twisted chain.  Note the similarity to the Skyrmion chain configurations 
obtained in \cite{HW08}.}\label{C4K0}
\end{figure}

\begin{figure}
\begin{minipage}{0.497\linewidth}
\centering
\vspace{-0.3cm}
\includegraphics[width=\linewidth]{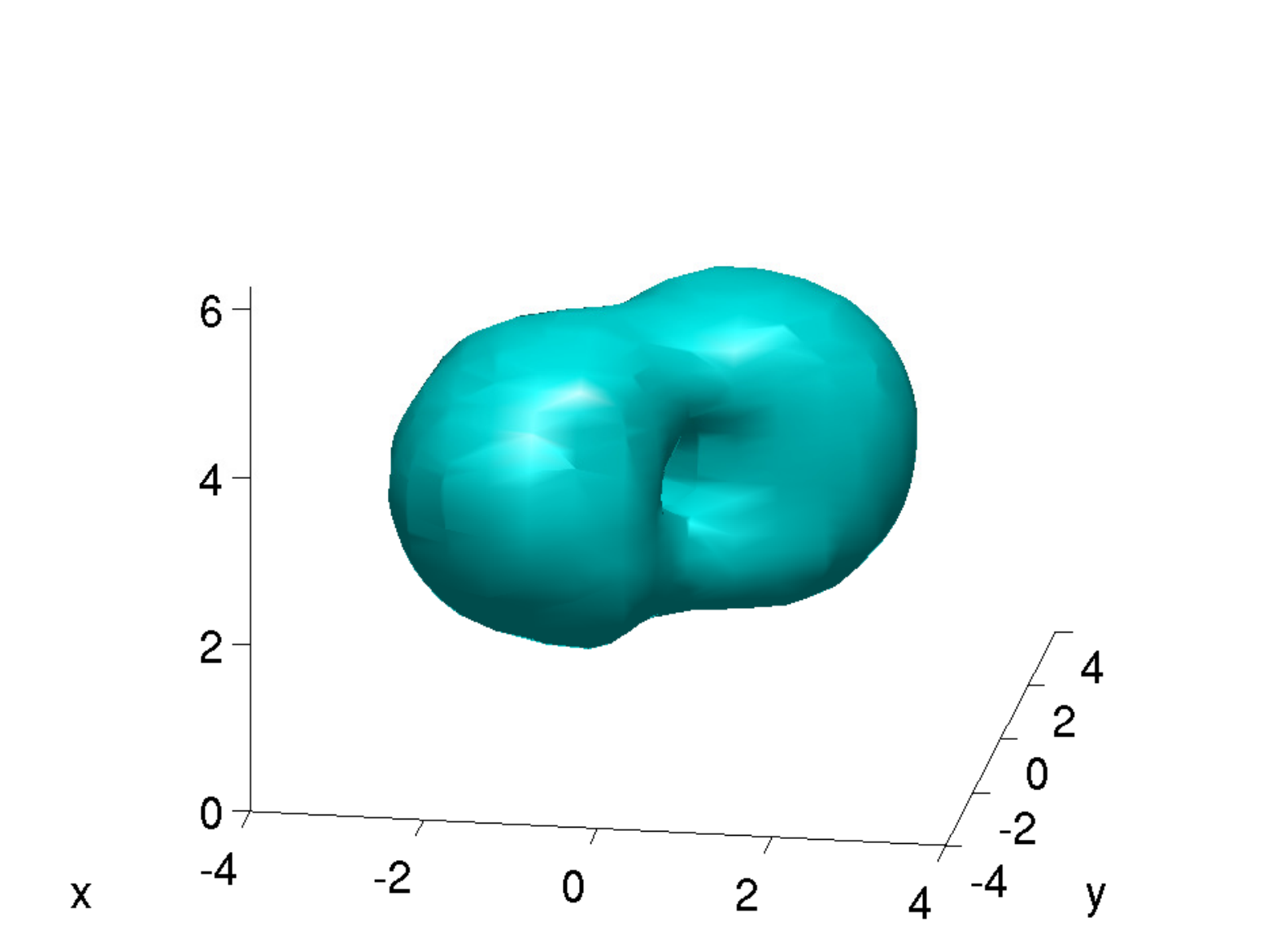}
\end{minipage}
\begin{minipage}{0.497\linewidth}
\centering
\includegraphics[width=\linewidth]{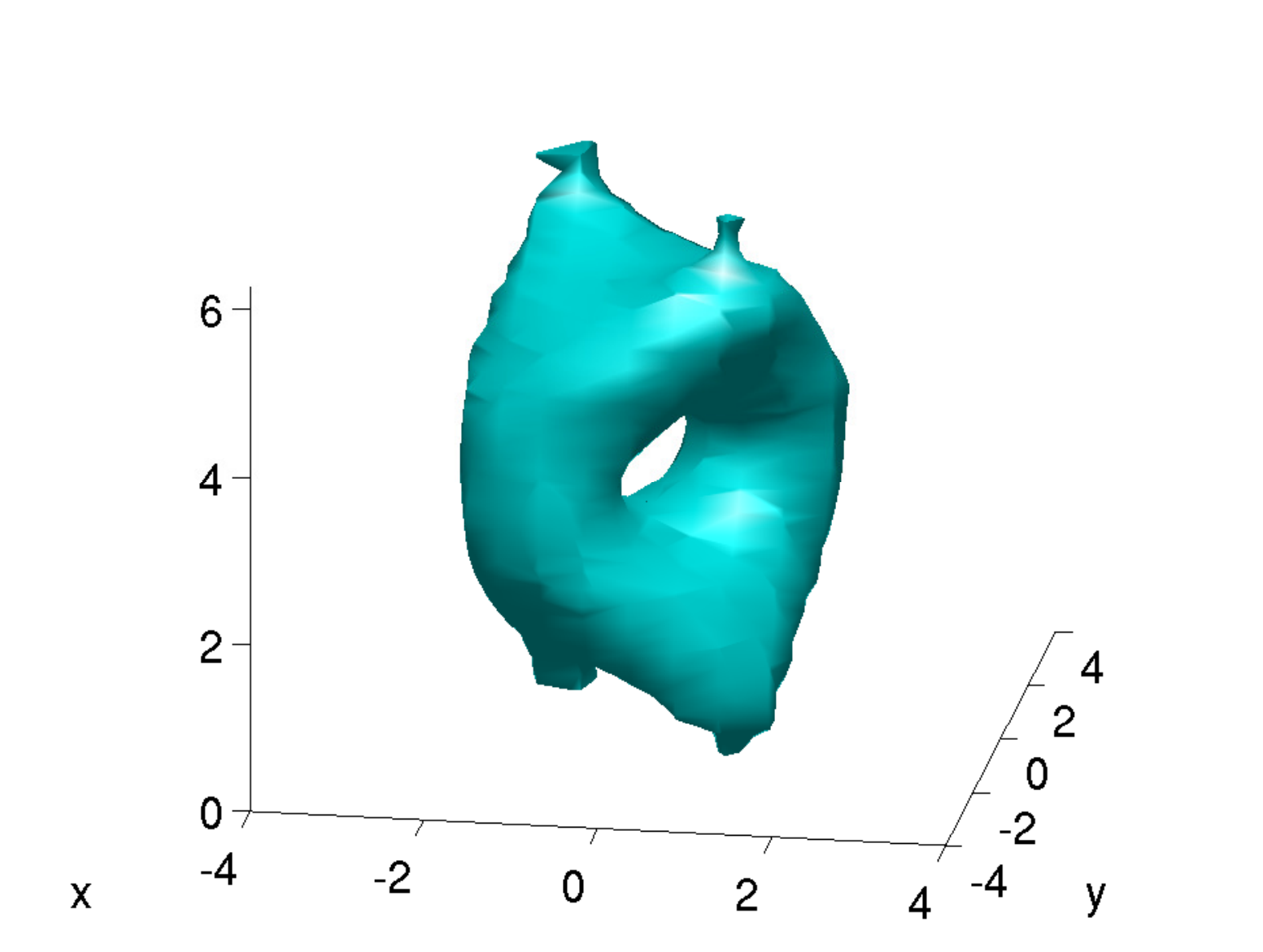}
\end{minipage}
\caption{Two points on the $W=\text{i}$ geodesic with $C=2$.  Left:~$W=2\text{i}$, right:~$W=1.125\text{i}$.  
As well as illustrating the scattering process, these energy density plots show how there is a transition 
between the $C=1$ case, where the energy is peaked in two regions near the $z$ axis, and the $C=4$ case, in 
which the energy is peaked away from the $z$ axis.  The `four pronged' structures of figure \ref{C1K0} can be 
visualised as splitting the right hand structure above along $z=\pi$.}\label{C2}
\end{figure}

 For $W=\text{i}$, the transition from small to large $C$ involves the resolution of the energy lumps of 
figure \ref{C1K0} into two constituents each.  Curiously, however, the constituents are not aligned with the 
$x$ and $y$ axes but with the lines $x\pm y=0$, such that the chain has been twisted by different amounts 
along its length, figure \ref{C4K0}.
The $W=\text{i}$ configuration on the $W\in\text{i}\mathbb{R}$ geodesic is an intermediate case between 
$W=\text{i}p$ with $p\gg1$ (for small $C$, this describes incoming chains along $x=y$ at $z=\beta/2$) and 
$0<p\ll 1$ (outgoing chains along $x=-y$ at $z=0$).  Figure \ref{C2} shows configurations with $W=2\text{i}$ 
and $W=1.125\text{i}$ for $C=2$, illustrating the fusing of two chains in the $W\in\text{i}\mathbb{R}$ 
geodesic and showing the transition between the $W=\text{i}$ chains of figures \ref{C1K0} and \ref{C4K0}.
\par
\begin{figure}
\begin{minipage}{0.485\linewidth}
\centering
\vspace{-0.3cm}
\includegraphics[width=0.95\linewidth]{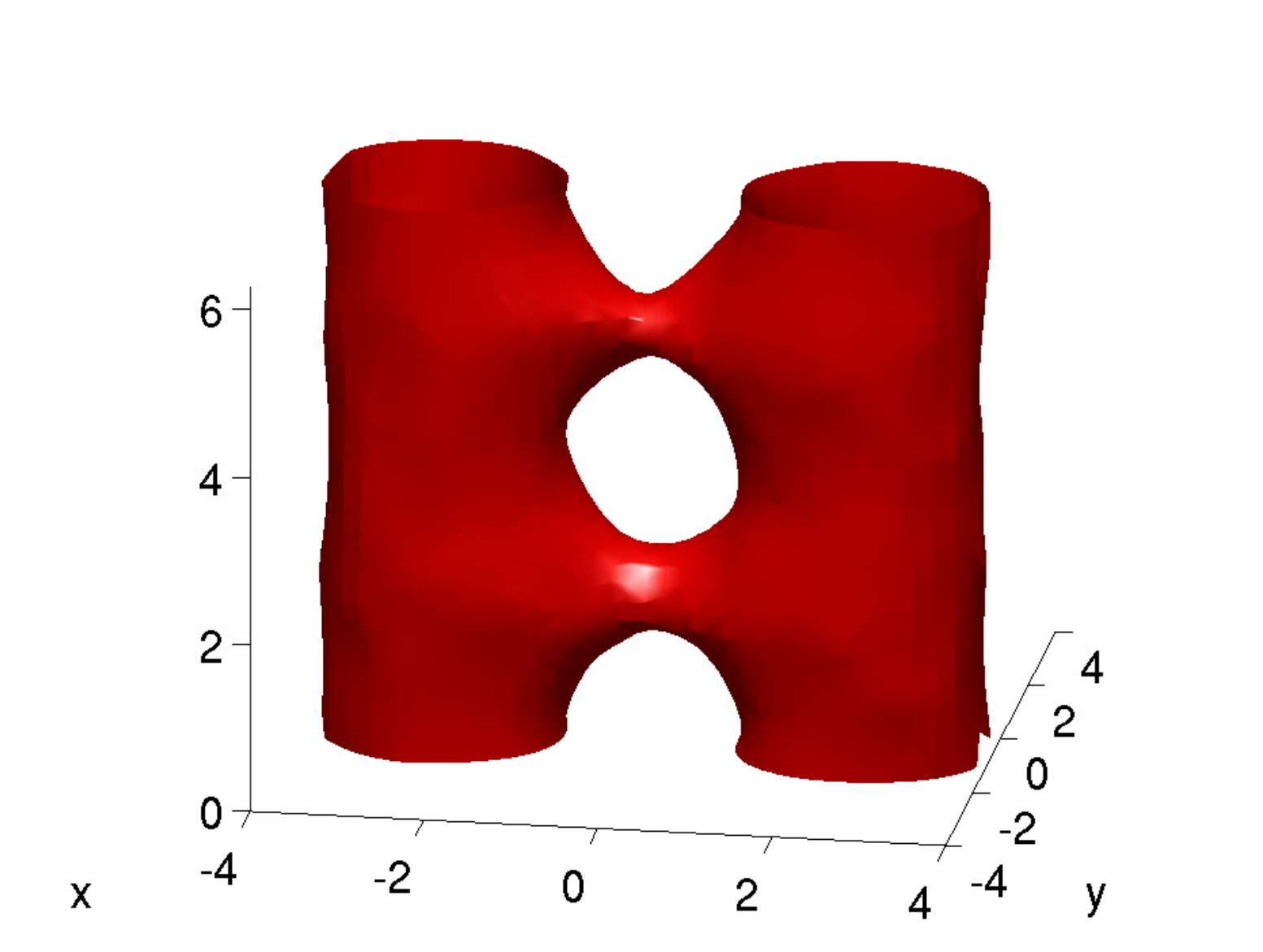}
\end{minipage}
\begin{minipage}{0.485\linewidth}
\centering
\includegraphics[width=0.95\linewidth]{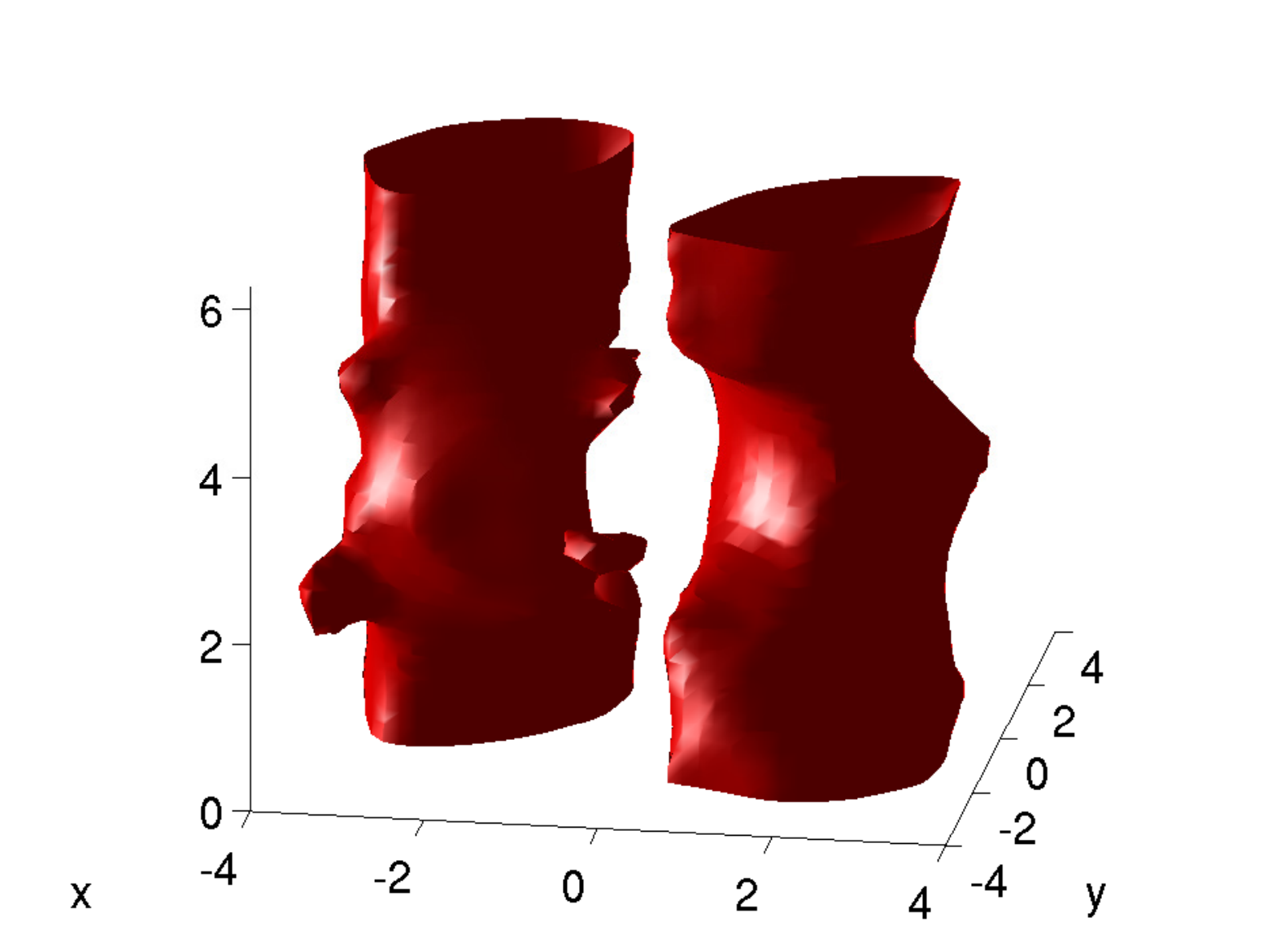}
\end{minipage}
\caption{Scattering for $C=4$.  Left: $W=1$, right: $W=\text{e}^{\text{i}\pi/3}$.  The $W=\text{i}$ 
configuration is shown in figure \ref{C4K0}.}\label{C4K12}
\end{figure}

 The $W=1$ configuration is the midpoint of scattering via the $W\in\mathbb{R}$ geodesic, for which 
outgoing chains are simply shifted by $\beta/2$ relative to the ingoing chains (figure \ref{C4K12} left).
\par We now describe the $|W|=1$ geodesic.  When $W=1$, we have a `chain of doughnuts', all equally aligned 
and with period $\beta/2$.  Altering the phase of $W$, the portion of the chain in the vicinity of 
$z=\beta/2$ rotates anticlockwise while that near $z=0$ and $z=\beta$ rotates clockwise such that at 
$W=\text{i}$ they are at $45^\circ$ to the coordinate axes.  These configurations are shown in 
figures \ref{C4K0} and \ref{C4K12} for $C=4$.
\par
\begin{figure}
\begin{minipage}{0.485\linewidth}
\centering
\vspace{-0.3cm}
\includegraphics[width=0.95\linewidth]{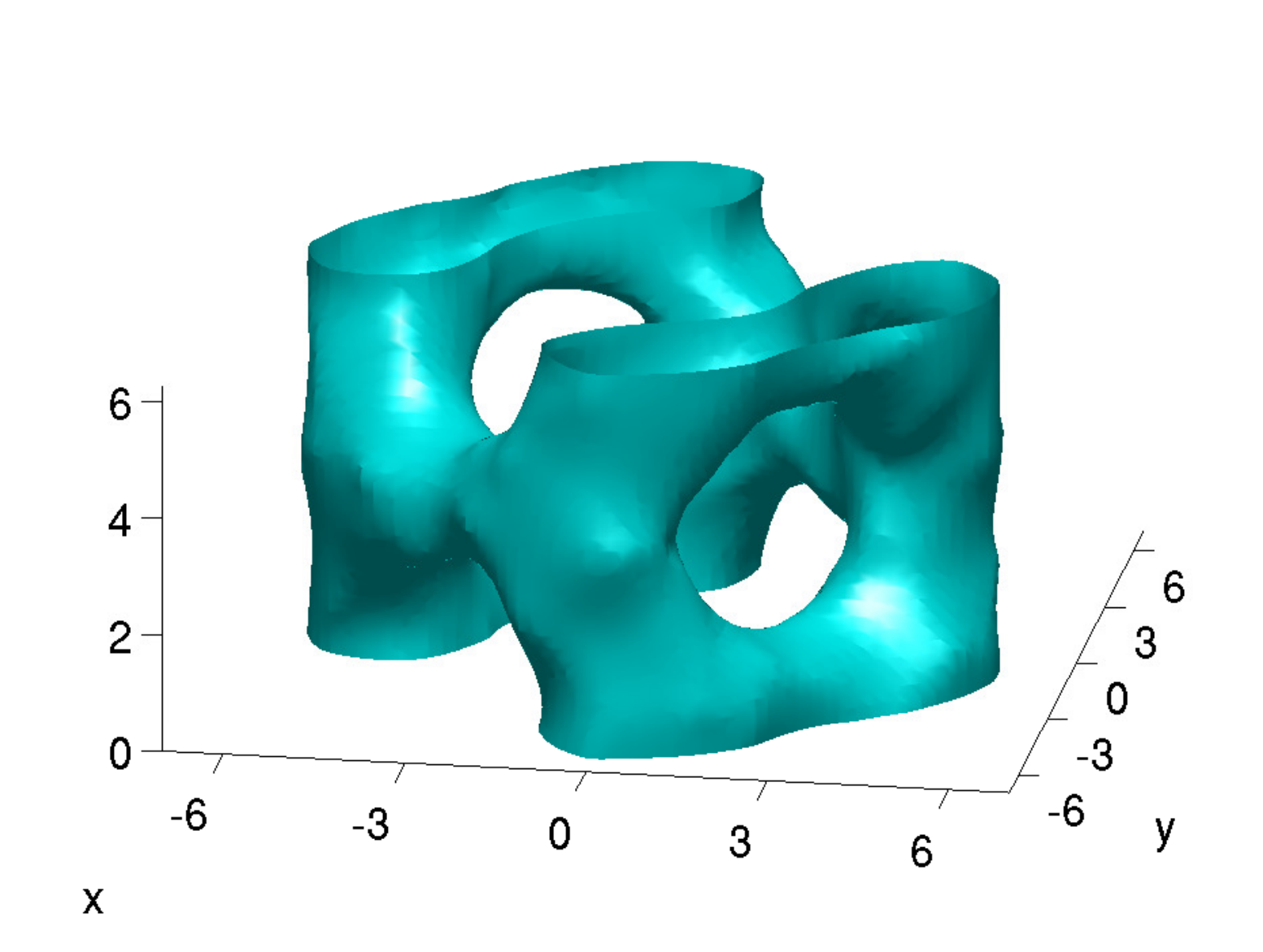}
\end{minipage}
\begin{minipage}{0.485\linewidth}
\centering
\includegraphics[width=0.95\linewidth]{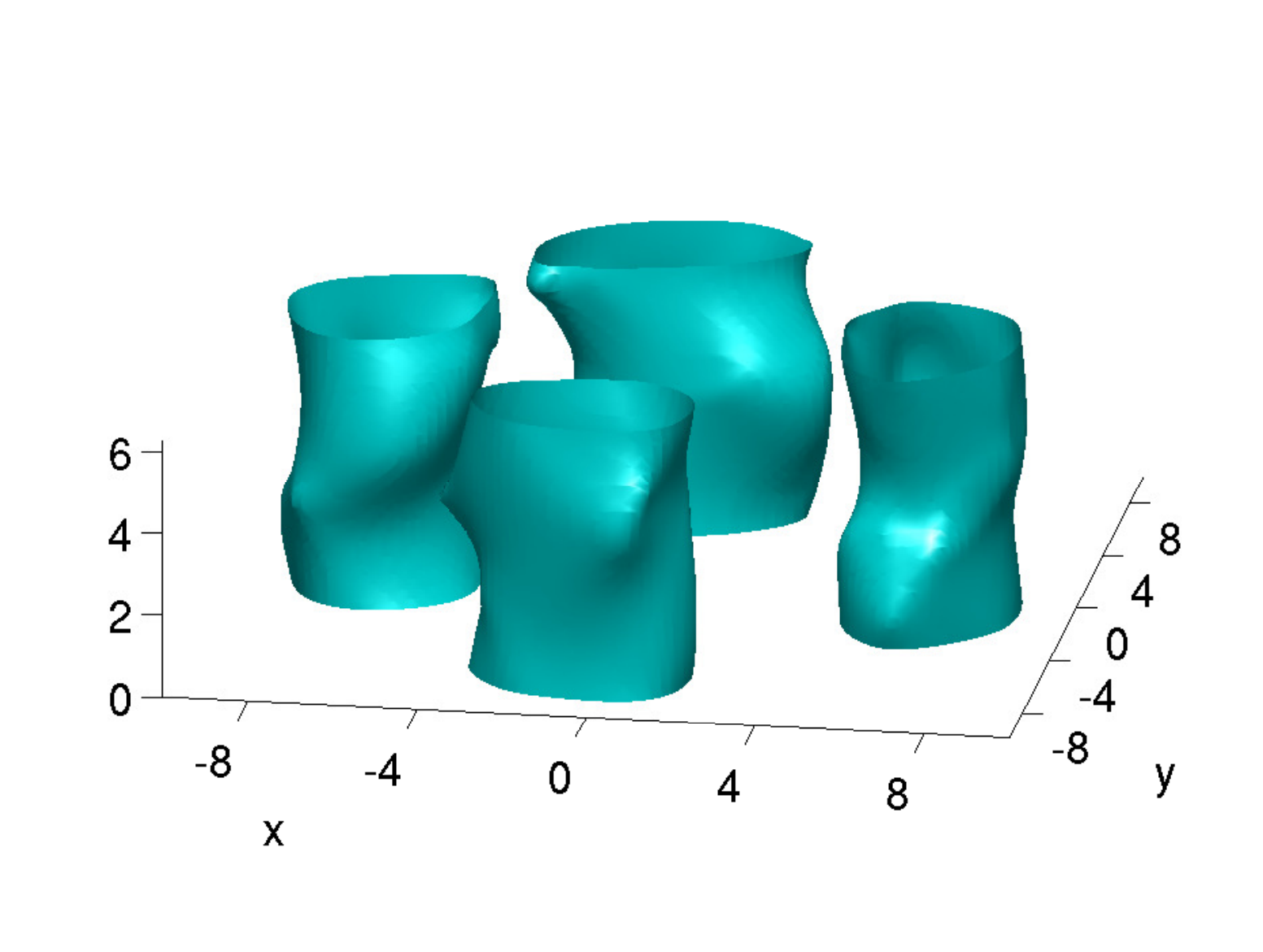}
\end{minipage}
\caption{Left: $C=16$, $W=\text{i}$, right: $C=36$, $W=\text{i}$.  Note how we approach the $z$-independent 
result of the spectral approximation (figure \ref{fig04a}, middle panel).}\label{C1636}
\end{figure}

 As $C$ is increased, the configuration deforms as shown in figure \ref{C1636}.  The energy lumps stretch 
in the $xy$ plane and fuse along $z$ such that when $C$ is large enough, there are tubes of energy density 
located in a cross shape aligned with the $x$ and $y$ axes.
\par
\begin{figure}
\begin{minipage}{0.485\linewidth}
\centering
\includegraphics[width=0.95\linewidth]{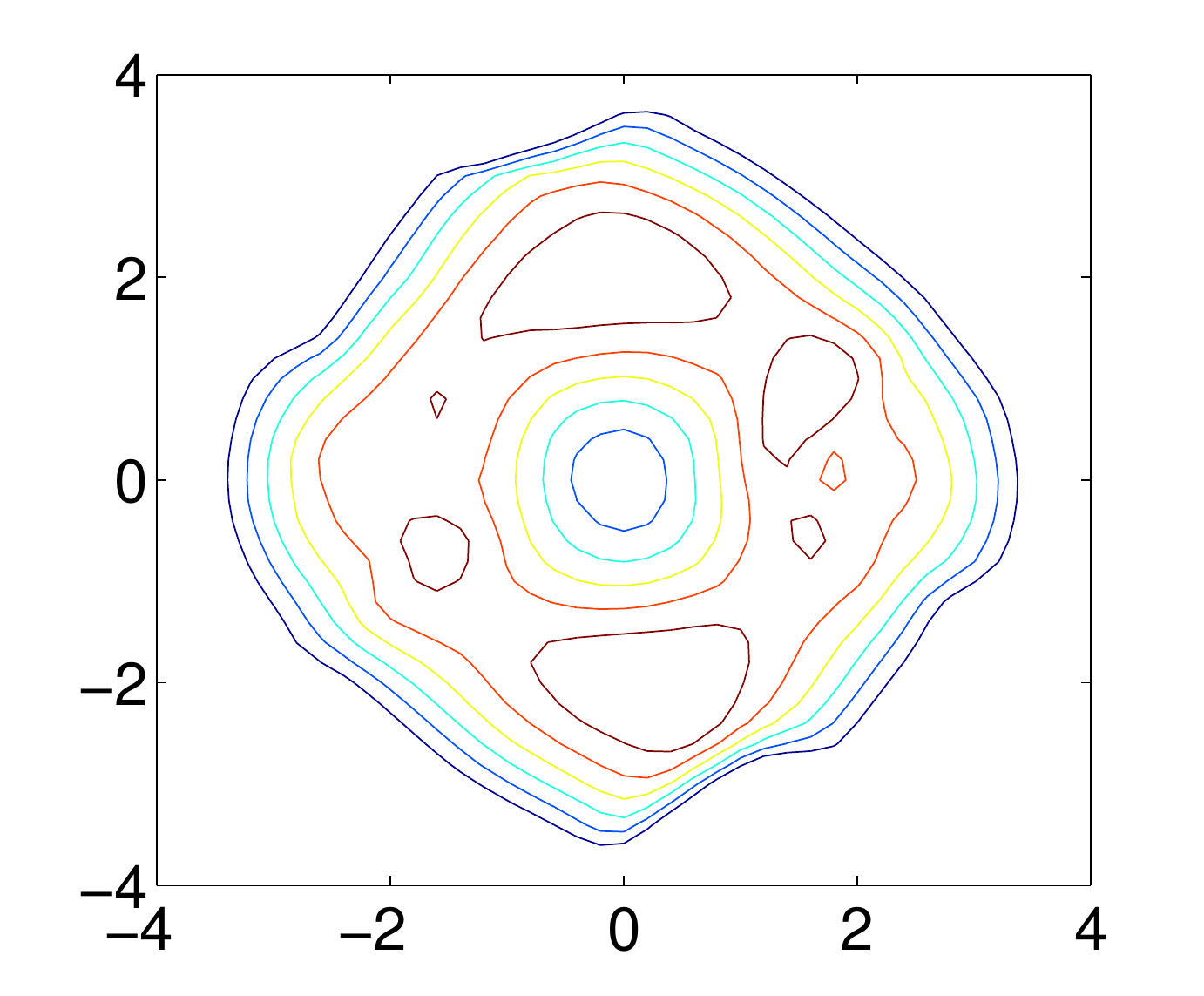}
\end{minipage}
\begin{minipage}{0.485\linewidth}
\centering
\includegraphics[width=0.88\linewidth]{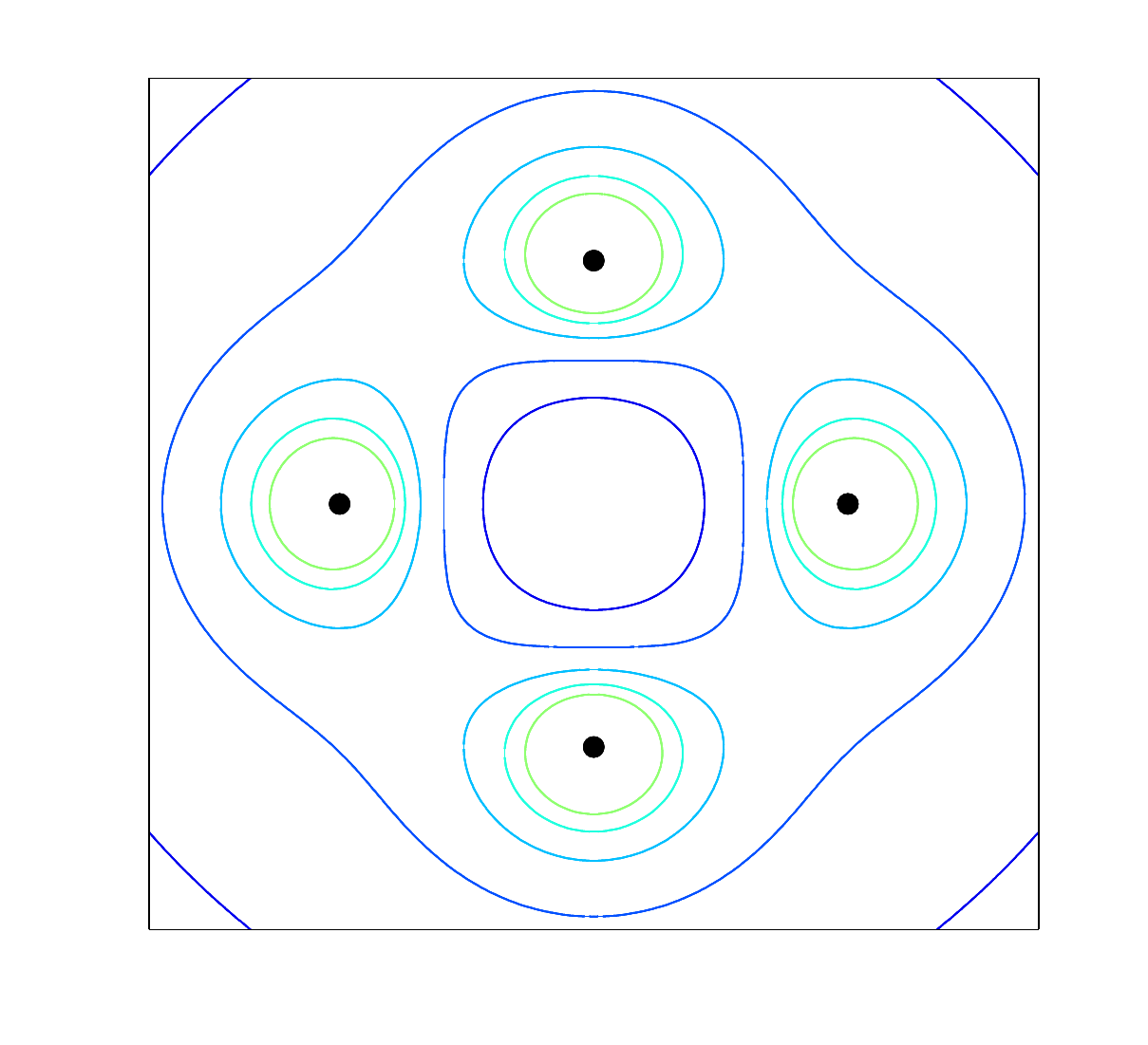}
\end{minipage}
\caption{Left: energy density integrated along a period for $C=4$, $K=0$.  Unlike in fig. \ref{C4K0} the 
energy peaks here are at the locations expected from the spectral approximation (right).  Note that this 
comparison requires a rescaling of the $x$, $y$ coordinates by a factor of $\sqrt{C}$.}\label{C4K0summed}
\end{figure}

 Even for intermediate values of $C$, one can make a link with the results of the spectral approximation 
by integrating the energy density along a $z$ period across the $xy$ plane.  The resulting quantity, shown in 
figure \ref{C4K0summed}, is found to resemble the energy density expected from the spectral approximation, 
insofar as the peaks are located along the coordinate axes and there is still a minimum at $x=y=0$.
\newpage
\section{Higher charge chains}\label{highercharges}
In this section we briefly make mention of the limits of varying $C$ with the charge $3$ periodic monopole 
described in \cite{Mal}.  For $K=0$ and small $C$ the charge $3$ configuration consists of an approximately 
toroidal charge $2$ monopole at $z=\beta/6$ and a roughly spherical charge $1$ monopole at $z=2\beta/3$.  For 
large $C$, the spectral approximation predicts a collection of tubes in a hexagonal configuration, with two of 
them found on the $x$ axis.
\par A numerical study of the effect of increasing $C$ is hindered by the need to make an appropriate choice 
of step size in performing the numerical derivative of $\hat{\Phi}$ to obtain the energy density via 
\eqref{endens}, as $\hat{\Phi}$ changes rapidly over small distances.  Nonetheless, the charge $2$ results 
suggest that for intermediate values of $C$ the configuration can be thought of as a chain of tetrahedra in 
alternating orientations disposed of as follows: place a base of the tetrahedron at $z=\beta/6$, oriented 
such that one of the vertices is aligned with the $y$ axis and another vertex is on the $z$ axis at 
$z=2\beta/3$.  The next tetrahedron shares the base, but is `upside down' (with a vertex at $z=-\beta/3$) and 
rotated around the $z$ axis by $\pi/3$ relative to the first.
\par One can understand higher charges in a similar way.  For instance, the charge $4$ configuration can 
occur in one of two types: either a chain of square pyramids disposed as with the tetrahedra of the charge 
$3$ chain (this is the `zeros apart' configuration of \cite{Mal}), or as a chain of cubes, each rotated by 
$\pi/4$ relative to the last (this is the `zeros opposite' configuration).


{\small\begin{thebibliography}{99}
\raggedright
\bibitem{ChK01}S.~Cherkis, A.~Kapustin, \emph{Nahm Transform for Periodic Monopoles and N=2 Super Yang-Mills Theory}, Commun.~Math.~Phys.~{\bf 218} (2001) 333, \href{http://arxiv.org/abs/hep-th/0006050}{\texttt{arXiv:hep-th/0006050}}
\bibitem{HW09}D.~Harland, R.~S.~Ward, \emph{Dynamics of periodic monopoles}, Phys.~Lett.~{\bf B 675} (2009) 262, \href{http://arxiv.org/abs/0901.4428}{\texttt{arXiv:0901.4428 [hep-th]}}
\bibitem{Mal13}R.~Maldonado, \emph{Periodic monopoles from spectral curves}, JHEP02(2013)099, \href{http://arxiv.org/abs/1212.4481}{\texttt{arXiv:1212.4481 [hep-th]}}
\bibitem{MW}R.~Maldonado, R.~S.~Ward, \emph{Geometry of periodic monopoles}, Phys.~Rev.~{\bf D 88} (2013) 125013, \href{http://arxiv.org/abs/1309.7013}{\texttt{arXiv:1309.7013 [hep-th]}}
\bibitem{LL98}K.~Lee, C.~Lu, SU(2)\emph{ calorons and magnetic monopoles}, Phys.~Rev.~{\bf D 58} (1998) 025011
\bibitem{War05}R.S.~Ward, \emph{Periodic monopoles}, Phys.~Lett.~{\bf B 619} (2005) 177, \href{http://arxiv.org/abs/hep-th/0505254}{\texttt{arXiv:hep-th/0505254}}
\bibitem{MS07}N.~Manton, P.~Sutcliffe, \emph{Topological Solitons}, CUP (2007)
\bibitem{ChK02}S.~A.~Cherkis, A.~Kapustin, \emph{Hyper-K\"ahler metrics from periodic monopoles}, Phys.~Rev.~{\bf D 65} (2002) 084015, \href{http://arxiv.org/abs/hep-th/0109141}{\texttt{arXiv:hep-th/0109141}}
\bibitem{Mal}R.~Maldonado, \emph{Higher charge periodic monopoles}, \href{http://arxiv.org/abs/1311.6354}{\texttt{arXiv:1311.6354 [hep-th]}}
\bibitem{BPP82}S.~A.~Brown, H.~Panagopoulos, M.~K.~Prasad, \emph{Two separated SU(2) Yang-Mills-Higgs monopoles in the Atiyah-Drinfeld-Hitchin-Manin-Nahm construction}, Phys.~Rev.~{\bf D 26} (1982) 854
\bibitem{Har}D.~Harland, private communication
\bibitem{Jar04}M.~Jardim, \emph{A survey on Nahm transform}, J.~Geom.~Phys.~{\bf 52} (2004) 313, \href{http://arxiv.org/abs/math/0309305}{\texttt{arXiv:math/0309305  [math.DG]}}
\bibitem{CG84}E.~Corrigan, P.~Goddard, \emph{Construction of Instanton and Monopole Solutions and Reciprocity}, Ann.~Phys.~{\bf 154}, 253-279 (1984)
\bibitem{HW08}D.~Harland, R.~S.~Ward, \emph{Chains of skyrmions}, JHEP12(2008)093, \href{http://arxiv.org/abs/0807.3870}{\texttt{arXiv:0807.3870 [hep-th]}}


\end{thebibliography}}
\end{document}